\documentclass[10pts,showpacs,preprint,aps]{revtex4}
\usepackage{graphicx}% Include figure files
\usepackage{dcolumn}% Align table columns on decimal point
\usepackage{bm}% bold math
\usepackage{amssymb}
\usepackage{amsmath}
\usepackage{color}
\begin{document}
\setcounter{page}{1}
\vskip 2cm
\title
{The origin of the mass of the Nambu-Goldstone bosons}
\author
{Ivan Arraut}
\affiliation{Department of Physics, Faculty of Science, Tokyo University of Science,
1-3, Kagurazaka, Shinjuku-ku, Tokyo 162-8601, Japan}

\begin{abstract} 
We explain the origin of the mass for the Nambu-Goldstone bosons when there is a chemical potential in the action which breaks explicitly the symmetry. The method is based on the number of independent histories for the interaction of the pair of Nambu-Goldstone bosons with the degenerate vacuum (triangle relations). The analysis suggests that under some circumstances, pairs of massive Nambu-Goldstone bosons can become a single degree of freedom with an effective mass defined by the superposition of the individual masses of each boson. Possible mass oscillations for the Nambu-Goldstone bosons are discussed.                                
\end{abstract}
\maketitle 

\section{Introduction}
The formulation of the Nambu-Goldstone theorem suggests the existence of certain number of gapless particles as a consequence of the existence of the same number of broken symmetries of the system under study \cite{Nambu3}. This was the brilliant observation of Yoichiro Nambu who was inspired in some fundamental problems inside the models of superconductivity in order to formulate what we know today as spontaneous symmetry breaking (SSB) \cite{Bardeen, Vala, Bolo, Nambu1, Nambu2}. Later this principle was translated to particle physics where it proved to be a powerful tool for explaining the electroweak unification, the associated BEH mechanism, as well as the chiral symmetry breaking \cite{Nambu3, Nambu4, Nambu5, Salam}. There are "exceptions" of the theorem where the number of Nambu-Goldstone bosons is lower than the number of broken generators \cite{Nambu55}. In addition, the dispersion relations for the Nambu-Goldstone bosons are not necessarily linear as it is expected from gapless particles \cite{Others}. This observation is summarized in the theorem of Nielsen and Chadha where a connection between the number of Nambu-Goldstone bosons and their dispersion was analyzed \cite{NC}. Then these previous observations were reviewed by some authors in order to generalize the counting rules for the Nambu-Goldstone bosons \cite{Murayama}. After then it was proved that the number of Nambu-Goldstone bosons as well as the associated dispersion relations are related to the number of independent histories representing the interaction of pairs of Nambu-Goldstone bosons with the degenerate vacuum, taken as a third particle with trivial phase. The number of independent histories are constrained by the Quantum Yang-Baxter equations (QYBE's) \cite{My papers, Integra}. In general, given the interaction of pairs of Nambu-Goldstone bosons with the degenerate vacuum (the degenerate vacuum taken as a fictitious particle with trivial phase), we can obtain a total of four histories of interaction. The QYBE's constraint the number of independent histories to two. The two independent histories will be related by a twist map (exchange of Nambu-Goldstone bosons), or equivalently by a time-reversal symmetry. It was demonstrated that when the number of independent histories is only one, then the dispersion relation of the Nambu-Goldstone bosons is quadratic and we only have one Nambu-Goldstone boson for a pair of broken symmetries. On the other hand, when the number of independent histories is two, then we have a linear dispersion relation and we have one Nambu-Goldstone boson for each broken symmetry \cite{My papers}. Previously, Nicolis and Piazza demonstrated that there is a gap associated to the Nambu-Goldstone bosons when there is a chemical potential associated to a finite charge density \cite{Nicolis}. The gap depends on the chemical potential as well as on the symmetry algebra of the broken generators. It comes out that in addition it is also necessary to make some arrangements to the counting of Nambu-Goldstone bosons in these situations \cite{Murayama}. However, it was not clear what is the effective gap when pairs of Nambu-Goldstone bosons with some specific value of mass (for each one), become a single degree of freedom. In this paper, by exploring the number of independent histories for the interaction of pairs of particles with the degenerate vacuum (triangle relations), we explain how under some circumstances pairs of Nambu-Goldstone bosons become effectively a single degree of freedom with an effective gap determined by the superposition of the individual masses for each Nambu-Goldstone boson.    


\section{General argument}  

\begin{figure}
	\centering
		\includegraphics[width=10cm, height=8cm]{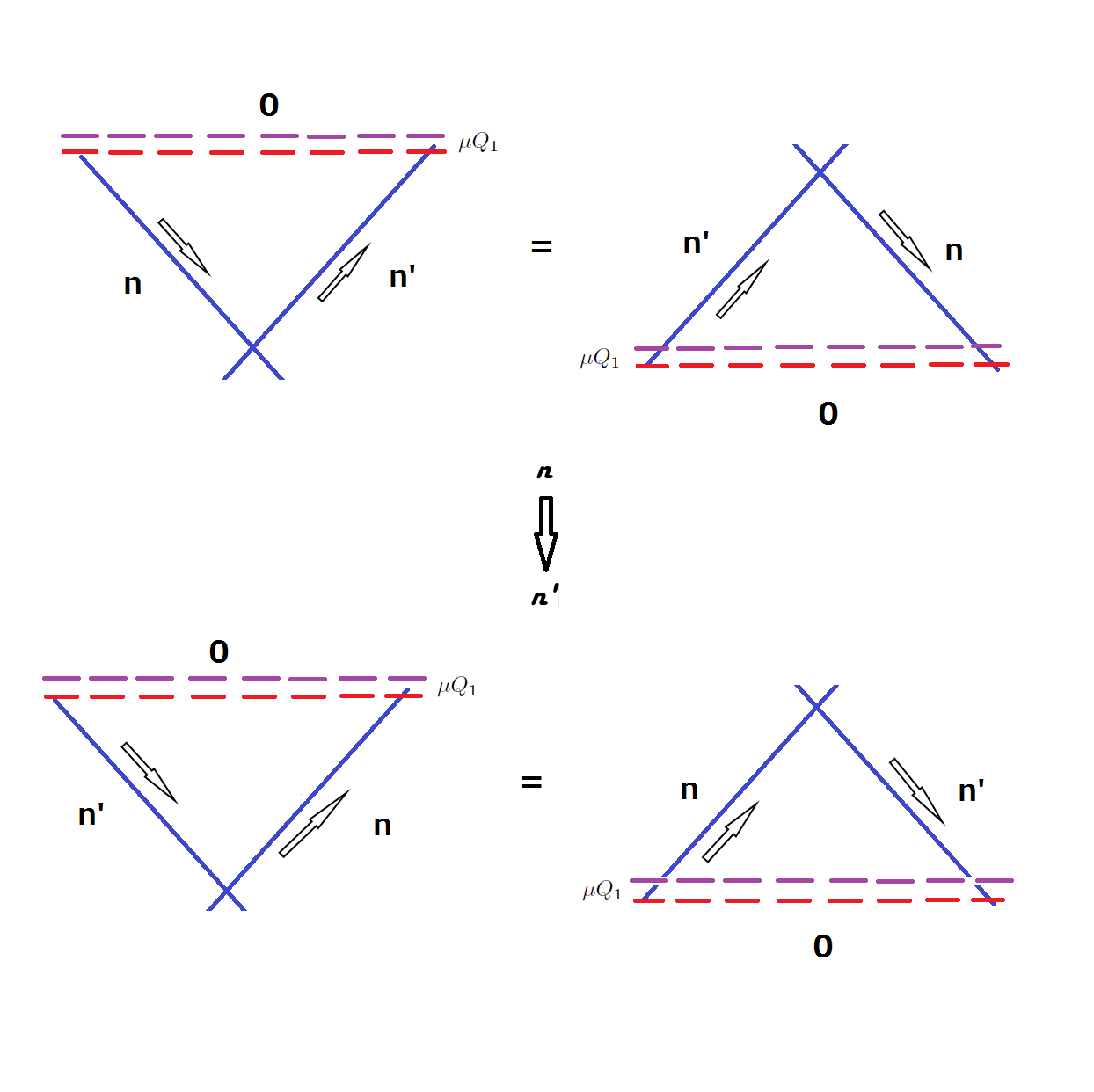}
	\caption{The Nambu-Goldstone bosons associated to the broken charge $Q_1$, meeting with the Nambu-Goldstone boson related to another broken generator of the system. The slopes measured with respect to the degenerate vacuum $0$ represent the spatial momentum of the particles. There are two degenerate vacuums with zero momentum. One of the vacuums also has zero energy but the other one (purple) has an energy shifted by $\mu Q_1$. The upper and lower figures are related by the exchange of particles $n\to n'$. The equalities respect the QYBE's.}
	\label{The titan3}
\end{figure}

\begin{figure}
	\centering
		\includegraphics[width=10cm, height=8cm]{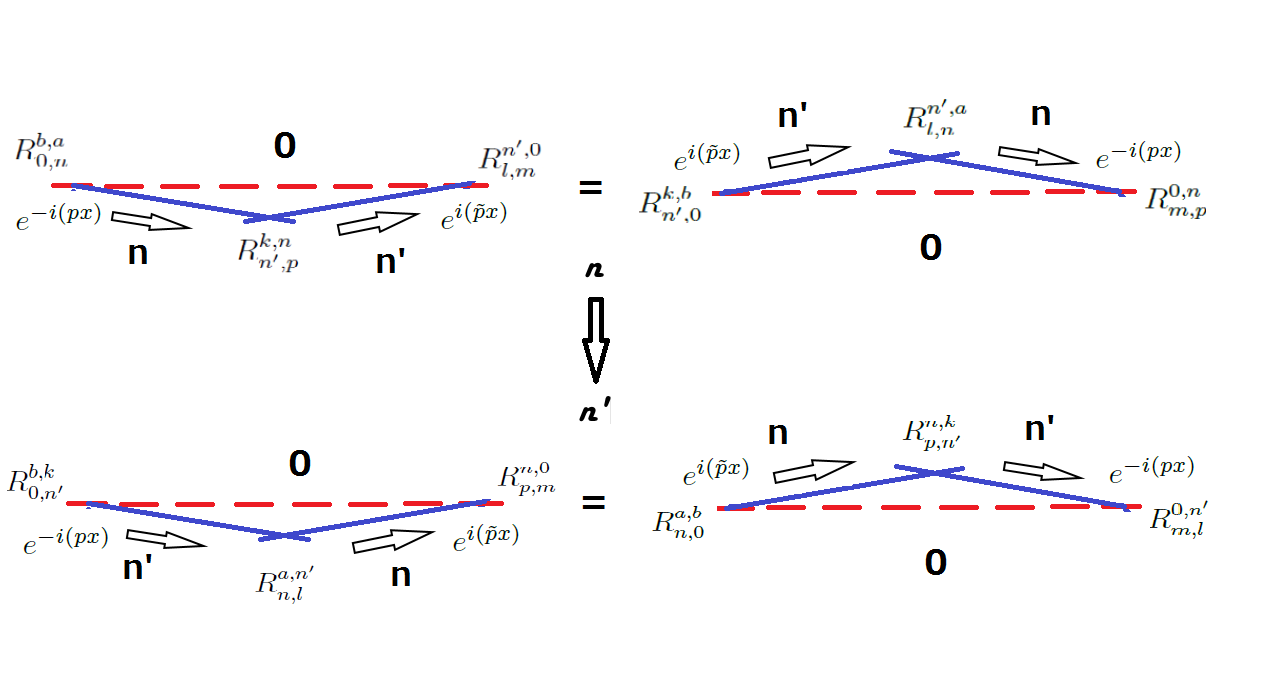}
	\caption{The infrared limit of the Nambu-Goldstone bosons. The triangles tend to be lines parallels to the degenerate vacuum. The figure illustrates the phase conventions used for the waves representing each particle. The vacuum is taken as a third particle of zero momentum.}
	\label{The titan4}
\end{figure}

We can analyze the interaction of a pair of Nambu-Goldstone bosons by understanding first how two plane waves meet at some region of spacetime. It is understood that the pair of Nambu-Goldstone bosons under analysis live over the degenerate vacuum and share a common region over the same vacuum. In addition we are working at the infrared limit such that the wave functions representing the pair of particles spread everywhere over the degenerate vacuum. The figure (\ref{The titan3}) illustrates the basic scenario of a pair of waves meeting in some region of spacetime. Initially we will take the particle (wave) denoted by $n$ as the Nambu-Goldstone boson corresponding to one of the broken generators different to the finite charge density term appearing in the action. We will then consider the particle $n'$ as the Nambu-Goldstone boson corresponding to the broken generator related to the charge appearing explicitly in the action. This case is interesting because it describes the origin of the mass of the Nambu-Goldstone bosons when the mechanism involved in their origin is based on symmetries. In other sections we generalize these results for explaining the cases where the pair of Nambu-Goldstone bosons under analysis are related to broken symmetries different to the one represented by the charge appearing explicitly in the action. In such a case, it will be already assumed that the pair of Nambu-Goldstone bosons interacting are already massive. We define two Hamiltonians $H_\mu$ and $H$ in a similar way as in ref. \cite{Nicolis}. The Hamiltonian $H$ satisfies the relation

\begin{equation}   \label{Qespo}
H\vert0_{SV}>=\mu Q_1\vert0_{SV}>,
\end{equation}
where $Q_1$ denotes the charge which appears explicitly in the action and it can be a broken generator. We can then define the Hamiltonian $H_\mu=H-\mu Q_1$ from which we define the ground state as $H_\mu\vert0_{SV}>=0$, consistent with the previous result. We can make the system to evolve in agreement with any Hamiltonian, namely, $H$ or $H_\mu$ as far as the results are consistent with each other. The only difference between the description of a system with respect to the Hamiltonian $H_\mu$ or the Hamiltonian $H$ is the time-direction considered for the evolution of the system. A system evolving in agreement with $H_\mu$, will not perceive explicitly the effects of the charge $Q_1$, but such effects will appear implicitly. On the other hand, a system evolving in agreement with $H$, will perceive explicitly the effects of the charge $Q_1$. At the end of the calculations, the physics obtained under both kind of evolutions is the same. In the same way, we can define the ground state with respect to any Hamiltonian, as far as we guarantee that our system is bounded from below. For example, for the ground state defined by $H_\mu$, there is no chance of having a particle with negative energy, which is correct in order to guarantee the stability of the theory. On the other hand, for a ground state defined in agreement with $H$, then $H\vert 0>=0$ and $H_\mu\vert0>=-\mu Q_1\vert0>$ for consistency. In this case, the particles should have positive energy with respect to $H$ and if for any reason they have a negative energy, it cannot be larger in magnitude than $\vert\mu q_a\vert$, with $q_a$ representing the eigenvalue of the operator $Q_1$. A particle with negative energy in this case should be interpreted as a particle with positive energy with respect to the real vacuum defined with respect to $H_\mu$. The Nambu-Goldstone bosons for the pair of particles will be represented by states evolving in agreement with the wave functions $e^{-i(px)}$ and $e^{-i(\tilde{p}x)}$, corresponding to the particle $n$ and $n'$ respectively. The phase convention is defined with respect to the vacuum. For the vacuum defined with respect to $H_\mu$, we will take $E_{\mu n}=0$ and ${\bf p}=0$, which gives the trivial wave function (phase) $e^{ip_\mu x}\to 1$. It is understood that $E_{\mu n}$ is the eigenvalue of the Hamiltonian $H_\mu$ and $E_n$ is the eigenvalue of the related Hamiltonian $H$. As has been said before, the vacuum with respect to $H_\mu$ is taken as a particle of trivial phase. However, we can define another degenerate vacuum with respect to the Hamiltonian $H$, which has a non-trivial phase represented as a particle with zero momentum but non-zero energy with the phase given by $e^{i\mu Q_1t}$ as it is shown by the purple dotted line in the figure (\ref{The titan3}).       
We have to define the phases for the particles with respect to the pair of degenerate vacuums. For the upper part of the figure (\ref{The titan3}), we will take $e^{-i(px)}=e^{-i(E_nt-{\bf p\cdot x})}$ as the function corresponding to the particle $n$. On the other hand, $e^{i(\tilde{p}x)}=e^{i(\tilde{E}_nt-{\bf \tilde{p}\cdot x})}$ will represent the function for the particle $n'$. The same convention applies for both sides of the equality. The equality is guaranteed by the QYBE's. For the lower figures, the phase convention is modified due to the exchange of particles $n\to n'$. In such a case, we take $e^{i(\tilde{p}x)}=e^{i(\tilde{E}_nt-{\bf \tilde{p}\cdot x})}=e^{i(\tilde{E}_nt+{\bf p\cdot x})}$ for the wave function corresponding to the particle $n$. In addition, we take $e^{-i(px)}=e^{-i(E_nt-{\bf p\cdot x})}=e^{-i(E_nt+{\bf \tilde{p}\cdot x})}$ for the plane wave representing the particle $n'$. Note the exchange of roles for the functions under the condition $n\to n'$. This convention is general, independent on whether the evolution of the particles is expressed with respect to $H_\mu$ or $H$. The figure (\ref{The titan4}) illustrates the phase conventions for the different particles. Note that it is the same figure (\ref{The titan3}) but at the infrared limit where the slopes of the lines representing the particles almost converge over the degenerate vacuum which is taken as a third particle with zero momentum. This is the interesting limit for the purposes of this analysis.          
\section{The sum of histories}

\begin{figure}
	\centering
		\includegraphics[width=10cm, height=8cm]{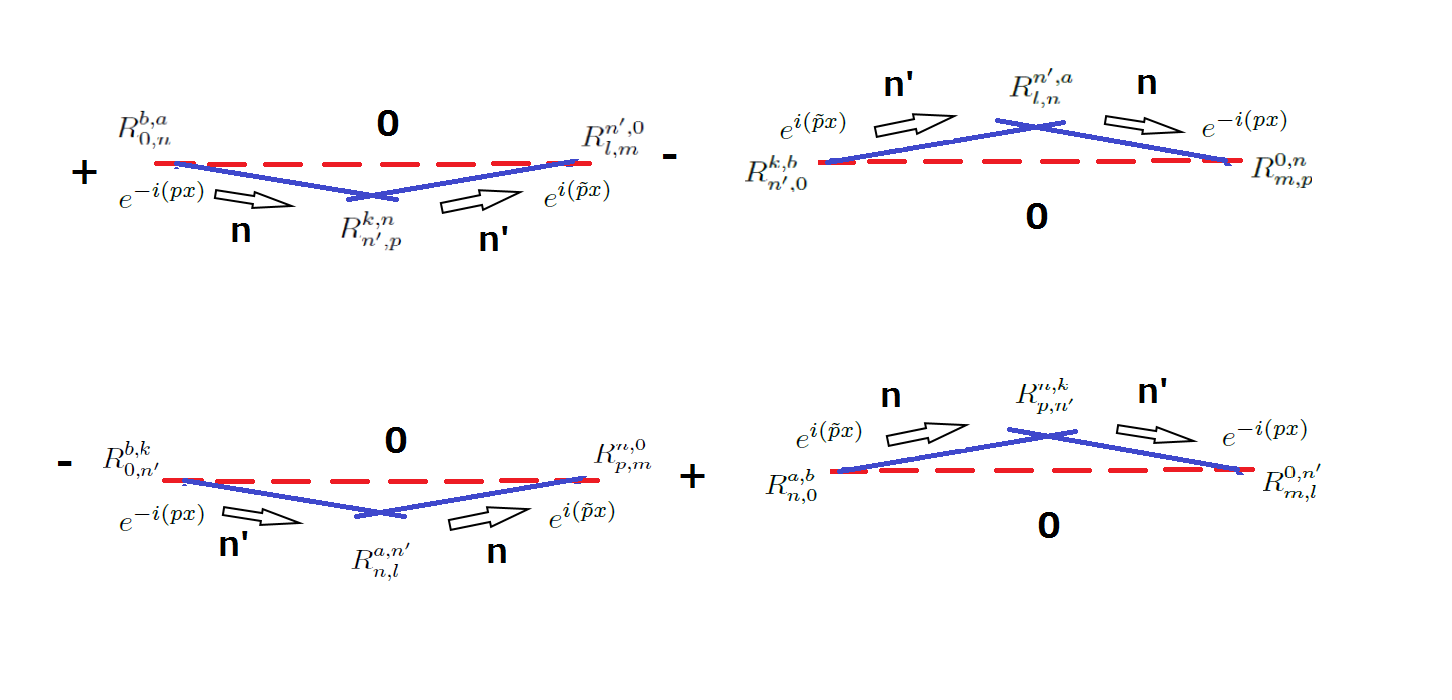}
	\caption{The sum of histories representing the interaction of the pairs of Nambu-Goldstone bosons and the degenerate vacuum taken as a third particle of zero momentum. An exchange $n\to n'$ is weighted with a minus factor. The same applies for an exchange of vacuum locations in the graph (from upper to lower side).}
	\label{The titan5}
\end{figure}

Each diagram showed in Fig. (\ref{The titan3}), illustrates a series of events (histories). When we sum the histories, then the exchange of location of the degenerate vacuum in the diagrams (from upper to lower) will be weighted with a minus sign. In addition, the exchange of particles represented by $n\to n'$ will be weighted by a minus sign. When there is a simultaneous exchange of both, vacuum and particles, then the graph will be weighted with a positive sign. The degenerate vacuum appearing in the diagrams, corresponds in reality to a third particle with trivial phase (zero energy and zero spatial momentum). Each graph in all the previous figures, corresponds to the product of three matrices, here denoted by $R$. In the product, the indices $n$, $n'$ and $0$ will appear as contractions between pairs of labels over which we have to sum. This is the case because the mentioned indices correspond to internal lines of the triangles. This is understood from the coordinate form of the QYBE's. Up to some phases to appear later, the sum illustrated in the figure (\ref{The titan5}) can be expressed as follows

\begin{eqnarray}   \label{spo}
\sum_{0, n, n'}<0_{DV}\vert Q_{l}(0)\vert n'><n'\vert Q_{p}(0)\vert n><n\vert \phi_{b}(x)\vert0_{DV}>(Ph_1)\nonumber\\
-<0_{DV}\vert \phi_{b}(x)\vert n'><n'\vert Q_{l}(0)\vert n><n\vert Q_{p}(0)\vert0_{DV}>(Ph_2)\nonumber\\
-<0_{DV}\vert Q_{p}(0)\vert n><n\vert Q_{l}(0)\vert n'><n'\vert \phi_{b}(x)\vert0_{DV}>(Ph_3)\nonumber\\
+<0_{DV}\vert \phi_{b}(x)\vert n><n\vert  Q_{p}(0)\vert n'><n'\vert Q_{l}(0)\vert0_{DV}>(Ph_4)=0,
\end{eqnarray}
where $Ph_i$ ($i=1,2,3,4$) corresponds to the phases for each term. They will be obtained through the spacetime invariance condition $Q_{p}(y)=e^{-ipy}Q_{p}(0)e^{ipy}$. In eq. (\ref{spo}), this condition is assumed. It is also assumed that the spatial translations are not spontaneously broken. If the charge $Q_1$ is a broken generator, then the symmetry under time translations, generated by the Hamiltonian $H$, will be also a broken symmetry if we define the ground state with respect to $H_\mu$. In this case, the time-translation symmetry is spontaneously broken. Equivalently, the symmetry under time translations generated by the Hamiltonian $H_\mu$ will be broken if we define the ground state with respect to $H$. Then the Lorentz symmetry is explicitly or spontaneously broken depending with respect to which Hamiltonian we are describing the physics of the system under study \cite{Nicolis}. We can introduce some auxiliary indices in eq. (\ref{spo}). They will represent the interactions of either, the pair of Nambu-Goldstone bosons $n$ and $n'$, or the interaction of one Nambu-Goldstone boson with the degenerate vacuum. Then eq. (\ref{spo}) can be written in terms of the product of three matrices as follows   

\begin{eqnarray}   \label{spo2}
R^{0,n'}_{\emph{\color{red}m},l}R^{n,\emph{\color{red}k}}_{p,n'}R^{\emph{\color{red}a},b}_{n,0}(Ph_1)-R^{n,0}_{p,\emph{\color{red}m}}R^{\emph{\color{red}a},n'}_{n,l}R^{b,\emph{\color{red}k}}_{0,n'}(Ph_2)-
R^{0,n}_{\emph{\color{red}m},p}R^{n',\emph{\color{red}a}}_{l,n}R^{\emph{\color{red}k},b}_{n',0}(Ph_3)+R^{n',0}_{l,\emph{\color{red}m}}R^{\emph{\color{red}k},n}_{n',p}R^{b,\emph{\color{red}a}}_{0,n}(Ph_4)=0.
\end{eqnarray}      
Here we have introduced the notation 

\begin{eqnarray}   \label{another form3}
R^{0,n'}_{\emph{\color{red}m},l}=<0_{DV}\vert Q_{\emph{\color{red}m},l}(0)\vert n'>,\;\;\;R^{n,\emph{\color{red}k}}_{p,n'}=<n'\vert Q_{\emph{\color{red}k},p}(0)\vert n>,\;\;\;\nonumber\\
R^{\emph{\color{red}a},b}_{n,0}=<n\vert \phi_{\emph{\color{red}a,}b}(x)\vert0_{DV}>,\;\;\;R^{n,0}_{p,\emph{\color{red}m}}=<n\vert Q_{p, \emph{\color{red}m}}(0)\vert0_{DV}>,\;\;\;\nonumber\\
R^{\emph{\color{red}a},n'}_{n,l}=<n'\vert Q_{l, \emph{\color{red}a}}(0) \vert n>,\;\;\;\;R^{b,\emph{\color{red}k}}_{0,n'}=<0_{DV}\vert \phi_{b, \emph{\color{red}k}}(x)\vert n'>.\;\;\;\;\nonumber\\
R^{0,n}_{\emph{\color{red}m},p}=<0_{DV}\vert Q_{\emph{\color{red}m},p}(0)\vert n>,\;\;\;\;R^{n',\emph{\color{red}a}}_{l,n}=<n\vert Q_{\emph{\color{red}a},l}(0)\vert n'>,\;\;\;\;\nonumber\\
R^{\emph{\color{red}k},b}_{n',0}=<n'\vert \phi_{\emph{\color{red}k},b}(x)\vert0_{DV}>,\;\;\;\;R^{n',0}_{l,\emph{\color{red}m}}=<n'\vert Q_{l, \emph{\color{red}m}}(0)\vert0_{DV}>,\;\;\;\;\nonumber\\
R^{\emph{\color{red}k},n}_{n',p}=<n\vert  Q_{p,\emph{\color{red}k}}(0)\vert n'>,\;\;\;\;R^{b,\emph{\color{red}a}}_{0,n}=<0_{DV}\vert \phi_{b,\emph{\color{red}a}}(x)\vert n>.\;\;\;\; 
\end{eqnarray}
Note the presence of the auxiliary indices marked with red color. They can be observed in the figure (\ref{The titan3}). Note that the pair of indices $\{p, \emph{\color{red}k}\}$ and $\{\emph{\color{red}a}, l\}$, represent interactions between the pair of Nambu-Goldstone bosons. Equivalently, the pair of indices $\{b, \emph{\color{red}k}\}$ and $\{\emph{\color{red}m}, l\}$, represent interactions between the degenerate vacuum and the particle $n'$ and finally the pair of indices $\{p, \emph{\color{red}m}\}$ and $\{\emph{\color{red}a}, b\}$, represent the interactions between the particles $n$ with the degenerate vacuum. After spatial integration, as well as taking into account time-independence, the terms representing the phases denoted by $Ph_i$ in eq. (\ref{another form3}), will tell us how the frequency and momentum go to zero simultaneously, telling us then the behavior of the dispersion relation for the effective Nambu-Goldstone bosons appearing at the end. Before going in further details, we will develop the concept of spontaneous symmetry breaking.
\section{Spontaneous symmetry breaking}

If we absorb the phases for each term inside the broken generators in eq. (\ref{spo}), and if in addition, we evaluate the expression in a single vacuum instead of summing over the degenerate vacuum, we will obtain the standard result

\begin{equation} \label{spo3}
<0_{SV}\vert [\phi_b(x),[Q_p(y),Q_l(z)]]\vert0_{SV}>\neq0. 
\end{equation} 
Note the presence of the spacetime dependence $y$ and $z$. This dependence will disappear after spatial integration and after imposing the standard time-independence condition. We want to remark at this point that eq. (\ref{spo3}) is just the SSB condition \cite{Nambu3}. Then we can conclude that the principle of SSB is a natural consequence of the interaction of pairs of particles with the degenerate vacuum after taking into account the appropriate weight factors (signs in front of the triangles). Then the SSB condition can be expressed in terms of the $R$-matrices obtained in eq. (\ref{spo2}), after introducing auxiliary indices and after summing over the degenerate vacuum. The degenerate vacuum is then a third particle with trivial momentum (zero slope). Note that the pair of indices $\{\emph{\color{red}m}, b\}$ live in the same space corresponding to the degenerate vacuum and they can be taken as equivalent. In addition, the pair of indices $\{\emph{\color{red}k}, l\}$ live in the same space corresponding to the particle $n'$. Finally, the pair of indices $\{\emph{\color{red}a}, p\}$ live in the space corresponding to the particle $n$. Then now the role of the auxiliary indices can be understood without any ambig\"uity. If a charge has the label $l$, then this broken generator is related to the Nambu-Goldstone boson $n'$. If a broken generator has the label $p$, then it is related to the Nambu-Goldstone boson $n$. Finally the order parameter with a label $b$ is always connected with the vacuum. By writing the SSB condition as in eq. (\ref{spo3}), we are assuming that the pair of broken generators $Q_p$ and $Q_l$ obey a Lie algebra. In our initial analysis, we will tale $Q_l=Q_1$. Then subsequently we will generalize the reault for any other pair of broken generators.  

\section{The dispersion relations and number of Nambu-Goldstone bosons}

The dispersion relations for the Nambu-Goldstone bosons, as well as their number, can be found under the assumption of spacetime translational invariance for the charges. In this section, we will only consider the case where one Nambu-Goldstone boson is related to the charge $Q_1$, and the other Nambu-Goldstone boson is related to a charge connected to any other symmetry spontaneously broken. As we have remarked before, we have two degenerate vacuums represented by a pair of parallel lines with zero slope but shifted with respect to each other by the gap $\mu q_a$, where $q_a$ is an eigenvalue of $Q_1$. The gap is illustrated in Fig. (\ref{The titan3}). In the figure, the red lines are the series of vacuums for which $H_\mu$ vanishes and the purple dotted lines are the vacuums under which $H$ vanishes. If the charge $Q_1$ is not a broken generator, then the gap between the lines disappears. The lines representing both degenerate vacuums, will be parallel because the spatial translations are not spontaneously broken and then the spatial momentum annihilates the vacuum independent on the Hamiltonian used for describing the evolution of the system.    In this section we will define the ground state (vacuum) as well as the evolution of the charges with respect to $H_\mu$. Then for $H_\mu$, we define the phases $e^{-ip_\mu x}\vert0>=e^{-i(E_\mu t-{\bf p\cdot x})}\vert0>=\vert0>$. Here $H_\mu\vert0>=E_\mu\vert0>=0$, which is consistent with what we have explained in eq. (\ref{Qespo}). On the other hand, since the Hamiltonian $H$ satisfies the relation (\ref{Qespo}), then the symmetry under time translations for the evolution of the system with respect to $H$ is broken due to its relations with the charge $Q_1$, as far as this charge is also a broken generator. Then the phase of this Hamiltonian with respect to the same ground state is defined in general by $e^{-ip x}\vert0>=e^{-i(Et-{\bf p\cdot x})}\vert0>=e^{i\mu Q_1t}\vert0>$. Here we will take the line $n'$ in the triangle relations illustrated in the figures (\ref{The titan3}), (\ref{The titan4}) and (\ref{The titan5}) as the line representing the Nambu-Goldstone bosons related to the broken charge $Q_1$. The slope of the line $n'$ with respect to the vacuum line ($0$), represents the spatial momentum of the Nambu-Goldstone bosons related to $Q_1$. By taking $Q_l(z)=e^{-i{pz}}Q_l(0)e^{ipz}=Q_1(z)$ and considering the evolution of the charges with respect to the Hamiltonian $H_\mu$, then the result (\ref{spo}), or equivalently, the result (\ref{spo3}) becomes

\begin{eqnarray}   \label{spo4}
\sum_{0, n, n'}<0_{DV}\vert Q_{1}(0)e^{i(\tilde{E}_{\mu n'}t-{\bf \tilde{p}_{n'}\cdot z})}\vert n'><n'\vert e^{-i(E_{\mu n'}t-{\bf p_{n'}\cdot y})}Q_{p}(0)e^{i(E_{\mu n}t-{\bf p_n\cdot y})}\vert n><n\vert \phi_{b}(x)\vert0_{DV}>\nonumber\\
-<0_{DV}\vert \phi_{b}(x)\vert n'><n'\vert e^{-i(\tilde{E}_{\mu n'}t-{\bf \tilde{p}_{n'}\cdot z})}Q_{1}(0) e^{i(\tilde{E}_{\mu n}t-{\bf \tilde{p}_{n}\cdot z)}} \vert n><n\vert e^{-i(E_{\mu n}t-{\bf p_n\cdot y})}Q_{p}(0)\vert0_{DV}>\nonumber\\
-<0_{DV}\vert Q_{p}(0)e^{i(E_{\mu n}t-{\bf p_n\cdot y})}\vert n><n\vert e^{-i(\tilde{E}_{\mu n}t-{\bf \tilde{p}_{n}\cdot z})}Q_{1}(0)e^{i(\tilde{E}_{\mu n'}t-{\bf \tilde{p}_{n'}\cdot z})}\vert n'><n'\vert \phi_{b}(x)\vert0_{DV}>\nonumber\\
+<0_{DV}\vert \phi_{b}(x)\vert n><n\vert e^{-i(E_{\mu n}t-{\bf p_n\cdot y})} Q_{p}(0) e^{i(E_{\mu n}t-{\bf p_n\cdot y})}\vert n'><n'\vert e^{-i(\tilde{E}_{\mu n'}t-{\bf \tilde{p}_{n'}\cdot z})} Q_{1}(0)\vert0_{DV}>=0.
\end{eqnarray}
The time-independence condition, together with the integration over the whole space, gives us the following results $E_{\mu n}=E_{\mu n'}$, $\tilde{E}_{\mu n}=\tilde{E}_{\mu n'}$, ${\bf p}_n={\bf p}_{n'}$ and ${\bf \tilde{p}_{n}}={\bf \tilde{p}_{n'}}$. Then we can ignore the phases related to the exponential terms having the combinations $E_{\mu n}-E_{\mu n'}$, $\tilde{E}_{\mu n}-\tilde{E}_{\mu n'}$, ${\bf p}_n-{\bf p}_{n'}$ and ${\bf \tilde{p}}_n-{\bf \tilde{p}}_{n'}$, understanding that the previous equalities are satisfied. If we introduce the notation given in eq. (\ref{another form3}) and then express eq. (\ref{spo4}) in terms of the $R$-matrices after introducing the auxiliary indices previously defined, we obtain

\begin{eqnarray}   \label{spo5}
R^{0,n'}_{\emph{\color{red}m},1}R^{n,\emph{\color{red}k}}_{p,n'}R^{\emph{\color{red}a},b}_{n,0}e^{i(\tilde{E}_{\mu n'}t-{\bf \tilde{p}_{n'}\cdot z})}-R^{n,0}_{p,\emph{\color{red}m}}R^{\emph{\color{red}a},n'}_{n,1}R^{b,\emph{\color{red}k}}_{0,n'}e^{-i(E_{\mu n}t-{\bf p_n\cdot y})}_{n'\to n}\nonumber\\
-R^{0,n}_{\emph{\color{red}m},p}R^{n',\emph{\color{red}a}}_{1,n}R^{\emph{\color{red}k},b}_{n',0}e^{i(E_{\mu n}t-{\bf p_n\cdot y})}+R^{n',0}_{1,\emph{\color{red}m}}R^{\emph{\color{red}k},n}_{n',p}R^{b,\emph{\color{red}a}}_{0,n}e^{-i(\tilde{E}_{\mu n'}t-{\bf \tilde{p}_{n'}\cdot z})}_{n\to n'}=0.\;\;\;\;\;
\end{eqnarray}      
Note that in the coordinate notation, the order of the matrices is irrelevant as far as the index contraction is the appropriate. The subindex $n\to n'$ below some phases, corresponds to the exchange of particles $n\to n'$. The final phases for the exchange of particles $n\to n'$ can be obtained naturally from the Yang-Baxter diagrams in the figures (\ref{The titan4}) and (\ref{The titan5}). They can be defined explicitly as $e^{-i(E_{\mu n}t-{\bf p_n\cdot y})}_{n'\to n}=e^{i(\tilde{E}_{\mu n'}t-{\bf \tilde{p}_{n'}\cdot y})}=e^{i(E_{\mu n}t+{\bf p_n\cdot y})}$ and $e^{-i(\tilde{E}_{\mu n'}t-{\bf \tilde{p}_{n'}\cdot z})}_{n\to n'}=e^{i(E_{\mu n}t-{\bf p_n\cdot z})}=e^{i(\tilde{E}_{\mu n'}t+{\bf \tilde{p}_{n'}\cdot z})}$. This result is general, independent on whether or not $n$ and $n'$ represent the same or different degrees of freedom, or equivalently, on whether or not the number of independent histories for the interaction of the particles is one or two. If we only have one independent history of interaction, then both degrees of freedom, namely, $n$ and $n'$ are the same and then $E_{\mu n}=\tilde{E}_{\mu n'}$. However, here ${\bf \tilde{p}_{\mu n'}}=-{\bf p_{\mu n}}$ due to the change of slope in the corresponding lines of the triangle relations. This condition over the spatial momentum is equivalent to say that the particles are approaching to each other. The two equalities considered in the figure (\ref{The titan3}), are a natural consequence of the QYBE's. However, in general, the pair of triangles of the upper part of the figure are different to the pair of triangles in the lowest part. The four triangles represent different histories for the interaction of the pair of particles and the degenerate vacuum. Then we can say that we have a total of four histories. However the QYBE's constraint the number of independent histories to only two. The two independent histories are related to each other by a twist map or time-reversal symmetry (exchange of particles $n\to n'$). If we have only one independent history, then this is only possible if $n=n'$, and then all the triangles of the figure are equivalent and we can then factorize all the terms with common coordinates. Then eq. (\ref{spo5}), becomes

\begin{eqnarray}
R^{0,n'}_{\emph{\color{red}m},1}R^{n,\emph{\color{red}k}}_{p,n'}R^{\emph{\color{red}a},b}_{n,0}\left(e^{i(\tilde{E}_{\mu n'}t-{\bf \tilde{p}_{n'}\cdot z})}+e^{i(\tilde{E}_{\mu n'}t+{\bf \tilde{p}_{n'}\cdot z})}\right)-R^{0,n}_{\emph{\color{red}m},p}R^{n',\emph{\color{red}a}}_{1,n}R^{\emph{\color{red}k},b}_{n',0}\left(e^{i(E_{\mu n}t-{\bf p_n\cdot y})}+e^{i(E_{\mu n}t+{\bf p_n\cdot y})}\right)
=0.\;\;\;\;\;
\end{eqnarray}        
This equation can be simplified as

\begin{eqnarray}   \label{timeindependent}
2R^{0,n'}_{\emph{\color{red}m},1}R^{n,\emph{\color{red}k}}_{p,n'}R^{\emph{\color{red}a},b}_{n,0}e^{i\tilde{E}_{\mu n'}t}cos\left({{\bf \tilde{p}_{n'}\cdot z}}\right)-2R^{0,n}_{\emph{\color{red}m},p}R^{n',\emph{\color{red}a}}_{1,n}R^{\emph{\color{red}k},b}_{n',0}e^{iE_{\mu n}t}cos({\bf p_n\cdot y})=0.\;\;\;\;\;
\end{eqnarray}
At this level it is clear that the momentum will go to zero quadratically and the frequency will vanish linearly. This is a proof of the fact that the condition $n=n'$ is equivalent to say that pairs of degrees of freedom originally with linear dispersion relation, become a single one with quadratic dispersion relation. Then the Nambu-Goldstone bosons related to the broken generators $Q_1$, are eaten up by the Nambu-Goldstone bosons related to the broken generator $Q_p$. This is an interesting phenomena and in some scenarios it can be interpreted as the fundamental origin of the Nambu-Goldstone's bosons mass. In other scenarios however, the gap of the Nambu-Goldstone bosons appears due to an ordinary contribution of the energy of the field associated to the charge $Q_1$. This is the situation in ferromagnetism and antiferromagnetism. Since the energy $E_{\mu n}$ is related to the Hamiltonian $H_\mu$, then the condition $E_{\mu n}=\tilde{E}_{\mu n'}\to 0$ is equivalent to the condition $E_n=\mu q_a$, where $q_a$ is an eigenvalue of $Q_1$. This gap is then generated dynamically due to the interaction of the pairs of degrees of freedom, which become a single one. The gap is defined by the relation $E_n-E_{\mu n}=\mu q_a$. Writing explicitly the results, from the time-independence condition of eq. (\ref{timeindependent}), as well as the spatial integration, we get

\begin{equation}
E_{\mu n}\to0\;\;\;\;E_n\to\mu q_a\;\;\;\;{\bf p}_n\to 0,
\end{equation}
with the dispersion relation

\begin{equation}
E_{\mu n}={\bf p}_n^2,
\end{equation}
which is equivalent to $E_n=\mu q_a+{\bf p}^2_n$. Note that this dependence is obtained by picking up the lowest order terms from the phases expansion in eq. (\ref{timeindependent}). The previous analysis, was based on the fact that the charges evolve with respect to the Hamiltonian $H_\mu$. Note that if $Q_1$ commutes with the other charges, then eq. (\ref{spo3}) becomes trivial. If we develop the same analysis with any other pair of charges different to $Q_1$ and satisfying eq. (\ref{spo3}), then the result will be equivalent but with a gap defined as the superposition of the individual gaps of the Nambu-Goldstone bosons interacting with each other.   

\section{Broken time translation symmetry}

When we analyzed the physics in the previous sections, we defined the vacuum with respect to the Hamiltonian $H_\mu$ (red dotted line in Fig. (\ref{The titan3})) and we also considered the evolution of the operators with respect to $H_\mu$. This means that the term $\mu Q_1$ did not appear explicitly although its physical effects still appear through the gap of the particles. The gap for the Nambu-Goldstone bosons appeared from the time-independent condition, obtaining then $E_{\mu n}\to 0$, but $E_n\to \mu q_a$. We will now analyze the evolution of the charges with respect to the Hamiltonian $H$ but still considering the vacuum with respect to $H_\mu$. Here we consider that the Hamiltonian $H$ is a broken generator since the charge $Q_1$ is also a broken generator. This is the only way for keeping consistency. Note that if we consider the vacuum with respect to $H$ (purple line in Fig. (\ref{The titan3})) such that $H\vert 0>=0$ (not broken), and if in addition, the charge $Q_1$ is not a broken generator, then we just recover the standard result where all the Nambu-Goldstone bosons are massless and independent (pairs do not converge into a single Goldstone boson). The physics described in this section is independent on which Hamiltonian is selected for describing the evolution of the charges and independent on the vacuum selected for describing the physics.  
If we define the vacuum with respect to $H$, then the frequencies defined by $E_n$ are gapless modes since $E_n\to0$. However, if in addition we have $Q_1$ as a broken generator, then it appears a negative gap for the frequencies $E_{\mu n}\to -\mu q_a$. This does not contradict the previous results and it should not be considered as an unphysical situation because what is important is to keep the condition $H_\mu\vert 0>=(H-\mu Q_1)\vert0>=0$, which is equivalent to $H\vert0>=(H_\mu+\mu Q_1)\vert0>=0$. Then having gapless frequencies for $E_n$, together with a negative gap for the frequency $E_{\mu n}=-\mu q_a$ is just equivalent to having a positive gap for $E_n=\mu q_a$ and a gapless condition for $E_{\mu n}$, proving then the consistency. Then whether a Nambu-Goldstone boson is massive or not is a matter of which vacuum is used for describing the physics of the system. Then the gap is correctly defined by the relative relation $E_n-E_{\mu n}=\mu q_a$ as before. In this section we consider the vacuum annihilated by $H_\mu$ (red line in the figure (\ref{The titan3})), but at the same time we consider the evolution of the operators with respect to $H$. In this case some interesting physical effects can be perceived. However, the results still keep consistency with the fact that there must be a relative gap of $\mu q_a$ between the frequencies $E_n$ and $E_{\mu n}$. We will analyze two possibilities inside this situation. Analogous results to the ones developed in this section, would be obtained if we consider the vacuum with respect to $H$ but the evolution of the charges with respect to $H_\mu$.  

\subsubsection{Pairs of Nambu-Goldstone bosons: $n'$ for the charge $Q_1$ and $n$ for another charge $Q_p$}   \label{great}

This is exactly the same case analyzed in detail previously. We define the vacuum with respect to $H_\mu$. However, since now we consider the evolution of the charges with respect to $H$, we have to make some changes. The charge $Q_1$ will evolve in the same way with respect to any Hamiltonian since $Q_1(t)=e^{-iHt}Q_1e^{iHt}=e^{-i(H_\mu+\mu Q_1)t}Q_1e^{i(H_\mu+\mu Q_1)t}=e^{-iH_\mu t}Q_1e^{iH_\mu t}$. This is the case because any operator commutes with itself and with any function depending on it. Then for this case we only have to focus on the operator $Q_p$. By finding the expression in terms of the Hamiltonian $H_\mu$, we get the result

\begin{eqnarray}   \label{spo41}
\sum_{0, n, n'}<0_{DV}\vert Q_{1}(0)e^{i(\tilde{E}_{\mu n'}t-{\bf \tilde{p}_{n'}\cdot z})}\vert n'><n'\vert e^{-i(E_{\mu n'}t-{\bf p_{n'}\cdot y})}e^{-i\mu Q_1t}Q_{p}(0)e^{i\mu Q_1t}e^{i(E_{\mu n}t-{\bf p_n\cdot y})}\times\nonumber\\
\vert n><n\vert \phi_{b}(x)\vert0_{DV}>-<0_{DV}\vert \phi_{b}(x)\vert n'><n'\vert e^{-i(\tilde{E}_{\mu n'}t-{\bf \tilde{p}_{n'}\cdot z})}Q_{1}(0)e^{i(\tilde{E}_{\mu n}t-{\bf \tilde{p}_{n}\cdot z)}}\vert n>\times\nonumber\\
<n\vert e^{-i(E_{\mu n}t-{\bf p_n\cdot y})}e^{-i\mu Q_1t}Q_{p}(0)e^{i\mu Q_1t}\vert0_{DV}>-<0_{DV}\vert e^{-i\mu Q_1t}Q_{p}(0)e^{i\mu Q_1t}\times\nonumber\\
e^{i(E_{\mu n}t-{\bf p_n\cdot y})}\vert n><n\vert e^{-i(\tilde{E}_{\mu n}t-{\bf \tilde{p}_{n}\cdot z})}Q_{1}(0)e^{i(\tilde{E}_{\mu n'}t-{\bf \tilde{p}_{n'}\cdot z})}\vert n'><n'\vert \phi_{b}(x)\vert0_{DV}>\nonumber\\
+<0_{DV}\vert \phi_{b}(x)\vert n><n\vert e^{-i(E_{\mu n}t-{\bf p_n\cdot y})} e^{-i\mu Q_1t}Q_{p}(0)e^{i\mu Q_1t}e^{i(E_{\mu n}t-{\bf p_n\cdot y})}\vert n'>\times\nonumber\\
<n'\vert e^{-i(\tilde{E}_{\mu n'}t-{\bf \tilde{p}_{n'}\cdot z})}Q_{1}(0)\vert0_{DV}>=0.
\end{eqnarray}        
Now we proceed to find the expression for $e^{-i\mu Q_1t}Q_{p}(0)e^{i\mu Q_1t}$, taking into account that $Q_p$ and $Q_1$ do not commute. The symmetries of the action obey the Lie algebra and then 

\begin{equation}   
[Q_p, Q_l]=if_{pl}^cQ_c.
\end{equation}
Here we consider the case $l=1$. With this commutator, we obtain \cite{Nicolis}

\begin{equation}   \label{MIX}
e^{-i\mu Q_1t}Q_{p}(0)e^{i\mu Q_1t}=(e^{f_1\mu t})_a^bQ_b.
\end{equation}
This result implies a mix of charges as the time evolve with respect to $H$. We consider the Hermitian basis where the structure constants are block diagonals matrices with the structure

\begin{equation}   \label{matrix} 
\begin{bmatrix}
       0 & -q_a      \\[0.3em]
       q_a & 0  \\[0.3em]
       \end{bmatrix}
\end{equation}
In this basis, the result (\ref{spo41}), expressed in terms of the $R$-matrices, becomes

\begin{eqnarray}   \label{spo51}
R^{0,n'}_{\emph{\color{red}m},1}R^{n,\emph{\color{red}k}}_{p,n'}R^{\emph{\color{red}a},b}_{n,0}e^{i\left((\tilde{E}_{\mu n'}-\mu q_a)t-{\bf \tilde{p}_{n'}\cdot z}\right)}-R^{n,0}_{p,\emph{\color{red}m}}R^{\emph{\color{red}a},n'}_{n,1}R^{b,\emph{\color{red}k}}_{0,n'}e^{-i\left((E_{\mu n}-\mu q_a)t-{\bf p_n\cdot y}\right)}_{n'\to n}\nonumber\\
-R^{0,n}_{\emph{\color{red}m},p}R^{n',\emph{\color{red}a}}_{1,n}R^{\emph{\color{red}k},b}_{n',0}e^{i\left((E_{\mu n}-\mu q_a)t-{\bf p_n\cdot y}\right)}+R^{n',0}_{1,\emph{\color{red}m}}R^{\emph{\color{red}k},n}_{n',p}R^{b,\emph{\color{red}a}}_{0,n}e^{-i\left((\tilde{E}_{\mu n'}-\mu q_a)t-{\bf \tilde{p}_{n'}\cdot z}\right)}_{n\to n'}=0.
\end{eqnarray}
Here we repeat the phase conventions explained previously such that $e^{-i\left((\tilde{E}_{\mu n'}-\mu q_a)t-{\bf \tilde{p}_{n'}\cdot z}\right)}_{n\to n'}=e^{i\left((E_{\mu n}-\mu q_a)t-{\bf p_{n}\cdot z}\right)}=e^{i\left((\tilde{E}_{\mu n'}-\mu q_a)t+{\bf \tilde{p}_{n'}\cdot z}\right)}$ and $e^{-i\left((E_{\mu n}-\mu q_a)t-{\bf p_n\cdot y}\right)}_{n'\to n}=e^{i\left((\tilde{E}_{\mu n'}-\mu q_a)t-{\bf \tilde{p}_{n'}\cdot y}\right)}=e^{i\left((E_{\mu n}-\mu q_a)t+{\bf p_n\cdot y}\right)}$. Here again the QYBE's constraint the number of independent histories to only two. Then after integration over the whole space and then taking the time-independence condition, here again the momentum goes to zero quadratically when the number of independent histories is reduced to one. Here again the Nambu-Goldstone bosons get a gap defined by $E_{\mu n}=\mu q_a$. For this case, the frequencies $E_n$ get a gap $E_n\to 2\mu q_a$. Then here again the relative value between the frequencies gap is $E_n-E_{\mu n}=\mu q_a$. Then the gap of the Nambu-Goldstone bosons is $\mu q_a$. The same relative gap will appear for any other case or combination of the previous cases. The mix of currents defined by the result (\ref{MIX}), is equivalent to a temporal exchange of the triangle lines representing the space where the Nambu-Goldstone bosons $n'$ live in the figures (\ref{The titan3}), (\ref{The titan4}) and (\ref{The titan5}). Then we have a mix of spaces or equivalently, a line becoming a different one after some temporal evolution. We can write explicitly how to get the gap from the previous expression. It is understood that the histories constrained by the QYBE's correspond to histories described by different spacetime coordinates ($y$ and $z$). Then we understand that the number of independent histories in the expression (\ref{spo51}) is two. When we reduce the number of histories to one, such that $n=n'$, then we reduce eq. (\ref{spo51}) to

\begin{eqnarray}   \label{spo51lala}
2R^{0,n'}_{\emph{\color{red}m},1}R^{n,\emph{\color{red}k}}_{p,n'}R^{\emph{\color{red}a},b}_{n,0}e^{i\left(\tilde{E}_{\mu n'}-\mu q_at\right)}cos({\bf \tilde{p}_{n'}\cdot z})-2R^{0,n}_{\emph{\color{red}m},p}R^{n',\emph{\color{red}a}}_{1,n}R^{\emph{\color{red}k},b}_{n',0}e^{i\left(E_{\mu n}-\mu q_a\right)t}cos({\bf p_n\cdot y})=0.
\end{eqnarray}
Then now we can see that at the lowest order in the expansions the momentum goes to zero quadratically and simultaneously the energy has a gap defined by $\mu q_a$. Then here we have a dispersion relation of the form

\begin{equation}
E_{\mu n}=\mu q_a+{\bf p}^2,
\end{equation}         
which can be considered as a quadratic dispersion relation. 
   
\subsubsection{Pairs of Nambu-Goldstone bosons: $n'$ for a charge $Q_l\neq Q_1$ and $n$ for another charge $Q_p\neq Q_1$}

For this case, the pair of charges under consideration have a non-trivial evolution with respect to $H$ when the vacuum is defined with respect to $H_\mu$. This case is just an extension of the one just analyzed and we will just write the final result, to be 

\begin{eqnarray}   \label{spo51nojoda}
R^{0,n'}_{\emph{\color{red}m},l}R^{n,\emph{\color{red}k}}_{p,n'}R^{\emph{\color{red}a},b}_{n,0}(e^{i\left((\tilde{E}_{\mu n'}-\mu q_a -\mu q_b)t-{\bf \tilde{p}_{n'}\cdot z}\right)})-R^{n,0}_{p,\emph{\color{red}m}}R^{\emph{\color{red}a},n'}_{n,l}R^{b,\emph{\color{red}k}}_{0,n'}(e^{-i\left((E_{\mu n}-\mu q_a-\mu q_b)t-{\bf p_n\cdot y}\right)}_{n'\to n})\nonumber\\
-R^{0,n}_{\emph{\color{red}m},p}R^{n',\emph{\color{red}a}}_{l,n}R^{\emph{\color{red}k},b}_{n',0}(e^{i\left((E_{\mu n}-\mu q_a-\mu q_b)t-{\bf p_n\cdot y}\right)})+R^{n',0}_{l,\emph{\color{red}m}}R^{\emph{\color{red}k},n}_{n',p}R^{b,\emph{\color{red}a}}_{0,n}(e^{-i\left((\tilde{E}_{\mu n'}-\mu q_a-\mu q_b)t-{\bf \tilde{p}_{n'}\cdot z}\right)}_{n\to n'})=0.
\end{eqnarray}
Since the vacuum is defined with respect to $H_\mu$, then the gap for $E_{\mu n}$ has to be positive definite. We ignore any possible negative gap contribution for $E_{\mu n}$ when the vacuum is defined with respect to $H_\mu$. This previous result can be simplified with the help of the QYBE's and by taking into account the phase convention previously explained, then we get

\begin{equation}   \label{spo51nojodadelajoda}
2R^{0,n'}_{\emph{\color{red}m},l}R^{n,\emph{\color{red}k}}_{p,n'}R^{\emph{\color{red}a},b}_{n,0}(e^{i(\tilde{E}_{\mu n'}-\mu q_a-\mu q_b)t})cos({\bf\tilde{p}_{n'}\cdot z})-2R^{n,0}_{p,\emph{\color{red}m}}R^{\emph{\color{red}a},n'}_{n,l}R^{b,\emph{\color{red}k}}_{0,n'}(e^{i(E_{\mu n}-\mu q_a-\mu q_b)t})cos({\bf p_{n}\cdot y})=0.
\end{equation}
Here the spatial momentum goes to zero quadratically as it is expected. Note that in this case, the gap has an interesting structure. The gap once again should be obtained from the time-independence condition and it is defined by $E_{\mu}=\mu q_a+\mu q_b$. The linear combination of charges appearing in the structure of the gap suggest that the pair of Nambu-Goldstone bosons were already massive before becoming a single degree of freedom. Then one Nambu-Goldstone boson with a gap $\mu q_a$ combines with another Nambu-Goldstone boson with gap $\mu q_b$, generating then effectively a single degree of freedom with an affective gap $\mu q_a+\mu q_b$ at the infrared level. Then at the moment when a pair of Nambu-Goldstone bosons become a single degree of freedom, there are two possibilities. The first one is that two massive Nambu-Goldstone bosons become effectively a single degree of freedom with a mass given by the superposition of the mass defined by each individual degree of freedom. The second option is one massless Nambu-Goldstone boson meets another one related to the charge $Q_1$, generating then a single degree of freedom, massive. This last scenario is illustrated by the analysis of the subsection (\ref{great}). It can be interpreted as a dynamical origin of the Nambu-Goldstone's bosons mass.         

\subsubsection{Possible future applications of the triangular formulation of the Nambu-Goldstone theorem}

The method developed in this paper can be perceived as a triangular formulation of the Nambu-Goldstone theorem. The systems analyzed in this paper have also been studied in the past by using other methods \cite{Nicolis}. However, in the near future we expect to extend the methods developed here in order to analyze some other systems where the physics is not yet completely understood. For example, since the formulation of this paper is based on the $R$-matrices, which in general are bi-linear objects; then the triangular approach becomes convenient (and natural) at the moment of analyzing cases where we consider order parameters formed as bi-linear objects. This is the case for the order parameter considered in the chiral condensation for example \cite{Chiral}, where the order parameter is given by \cite{Chiral}

\begin{equation}
<\bar{\psi}^a\psi^b>=R^{a,b}_{c, d}\epsilon^c\otimes\epsilon^d,
\end{equation}        
which is evidently a bi-linear object corresponding to the $R$-matrices constructed in this paper. Note that for the cases analyzed here, we focused on systems where the order parameter might correspond to a scalar or a vector. The same applies for the corresponding symmetries involved. However, it is well-known that scalar and vectors are just zero and first order rank tensors. Then the $R$-matrices can be also perfectly adapted for such situations if we understand the role taken by each index as has been done in this paper. In other words, the triangular formulation proposed here is general and can be perfectly extended to the analysis of different situations, independent on whether or not the order parameter and the symmetries of the systems correspond to bi-linear objects. In addition, the formalism  can be also applied to situations where the perturbative approaches to Quantum Field Theory (QFT) cannot be used or when the structure of the vacuum seems to be complicated to analyze by using conventional methods. This is the case for example of the quark-gluon plasma, which is in general difficult to analyze by using conventional methods. The formalism can also be extended for the cases where we can consider finite temperature corrections. With this formulation, we expect in general to analyze some difficult problems inside the scenario of Quantum Chronodynamics (QCD). These treatments, as well as other physical situations, will be considered in coming papers.      

\section{Conclusions}

In this paper we have explained the different mechanisms under which the Nambu-Goldstone bosons can get a mass when there is a chemical potential, breaking then the symmetry under time-translations for one of the Hamiltonians describing the physics of the system. We used the method based on expressing the spontaneous symmetry breaking condition as a sum of histories. In total for the interaction of a pair of Nambu-Goldstone bosons with the degenerate vacuum taken as a third particle of zero momentum, we have four histories of interaction. However, the QYBE's constraint the number of independent histories to two. When the number of independent histories is only one, then the pair of Nambu-Goldstone bosons become a single degree of freedom with quadratic dispersion relation. This triangular formulation suggests that in some circumstances, some Nambu-Goldstone bosons are eaten up by the partners, generating then the gap dynamically. The gap and the physics described is independent on the definition of the vacuum and independent on the Hamiltonian selected in order to describe the evolution of the system. This result evidently matches with what was found by other authors in \cite{Nicolis}. The structure for the gaps becomes richer when we consider the pair of charges in the triangle relations as different to $Q_1$, with the vacuum defined with respect to $H_\mu$ and the evolution of the system with respect to $H$. In this case, the spectrum of masses of the Nambu-Goldstone bosons appear explicitly. The analysis suggests that for this case a pair of massive Nambu-Goldstone bosons with linear dispersion relation become a single degree of freedom with quadratic dispersion relation and with the mass determined by the sum of the individual gaps of the original degrees of freedom. We have to remark at this point that the previous methods for analyzing the mechanism of spontaneous symmetry breaking are good enough for confronting the kind of problems described in this paper. The purpose of this paper is to introduce a new tool for analyzing the spontaneous symmetry breaking phenomena. We have proved then that the new method developed here can also predict the same physics which was previously known, which is the minimal requisite in order to be valid. This new method will become fundamentally important when we confront some difficult problems such as those related to the quark-gluon plasma and problems related to the quark-confinement in general, together with the associated chiral symmetry breaking as has been mentioned in the last section. In this paper we have just mentioned some possible advantages of the method but not developed them in deep detail. Such deeper analysis related to QCD and its vacuum structure will correspond to a coming paper.  
 \\\\

{\bf Acknowledgement}
I. A is supported by the JSPS Post-doctoral fellowship for oversea Researchers.

\end{document}